\documentclass[conference]{IEEEtran}

\usepackage{cite}
\usepackage{graphicx,color,epsfig,rotating}
\usepackage{amsfonts,amsmath,amssymb,bbm}
\usepackage{algorithm, algorithmic}
\usepackage{subfigure}
\usepackage{amsmath}
\usepackage{cite}
\usepackage{mdwtab}
\usepackage{subfigure}
\usepackage{placeins}
\usepackage{psfrag, graphicx}
\usepackage[latin1]{inputenc}
\usepackage{amssymb}
\usepackage{makeidx}
\usepackage{url}
\usepackage{epstopdf}

\long\def\comment#1{}

\newfont{\bbb}{msbm10 scaled 700}

\newfont{\bb}{msbm10 scaled 1100}

\newcommand{\Kc}{{\cal K}}

\newcommand{\eqdef}{\stackrel{\Delta}{=}}

\newcommand{\be}{\begin{equation}}
\newcommand{\ee}{\end{equation}}
\newcommand{\bea}{\begin{eqnarray}}
\newcommand{\eea}{\end{eqnarray}}

\newtheorem{theorem}{Theorem}%[section]
\newtheorem{corollary}{Corollary}%[section]
\begin{document}

\title{A New Combinatorial Design of Coded Distributed Computing}

%\author{Nicholas Woolsey, Rong-Rong Chen and Mingyue Ji
%\thanks{The authors are with the Department of Electrical Engineering,
%University of Utah, Salt Lake City, UT 84112, USA. (e-mail: nicholas.woolsey@utah.edu, rchen@ece.utah.edu and mingyue.ji@utah.edu)}
%}

\author{
    \IEEEauthorblockN{ Nicholas Woolsey,
		Rong-Rong Chen, and Mingyue Ji }
	\IEEEauthorblockA{Department of Electrical and Computer Engineering, University of Utah\\
		Salt Lake City, UT, USA\\
		Email: \{nicholas.woolsey@utah.edu,
		 rchen@ece.utah.edu,
		mingyue.ji@utah.edu\}}

}

\maketitle

\thispagestyle{empty}
\pagestyle{empty}

\vspace{-0.5cm}

\begin{abstract}
Coded distributed computing introduced by Li et al. in 2015 is an efficient approach to trade computing power to reduce the communication load in general distributed computing frameworks such as MapReduce. In particular, Li et al. show that increasing the computation load in the Map phase by a factor of $r$ can create coded multicasting opportunities to reduce the communication load in the Reduce phase by the same factor. However, there are two major limitations in practice. First, it requires an exponentially large number of input files (data batches)
when the number of computing nodes gets large. Second, it forces every $s$ computing nodes to compute one Map function, which leads to a large number of Map functions required to achieve the promised gain. In this paper, we make an attempt to overcome these two limitations by proposing a novel coded distributed computing approach based on a combinatorial design. We demonstrate that when the number of computing nodes becomes large, 1) the proposed approach requires an exponentially less number of input files; 2) the required number of Map functions is also reduced exponentially. Meanwhile, the resulting computation-communication trade-off maintains the multiplicative gain compared to conventional uncoded unicast and achieves the information theoretic lower bound asymmetrically for some system parameters.
\end{abstract}

%\begin{IEEEkeywords}
%Wireless Distributed Computing
%\end{IEEEkeywords}

%\newpage

\section{Introduction}
\label{section: intro}

Coded distributed computing introduced in \cite{li2018fundamental} is an efficient approach to reduce the communication load in the distributed computing framework such as MapReduce \cite{dean2008mapreduce}. %and Spark \cite{zaharia2010spark}. %and mitigate the impact of stragglers in distributed computing systems.
In this type of distributed computing networks, in order to compute the output functions, the computation is decomposed into ``Map" and ``Reduce" phases. First, each computing node computes intermediate values using the local input data files according to the designed Map functions, then computed intermediate values are exchanged among the computing nodes in order to obtain the final output functions for each node using their designed Reduce functions. The operation of exchanging intermediate values is called ``data shuffling" or ``Shuffle phase", which appears to limit the performance of distributed computing applications due to the amount of transmitted traffic load \cite{li2018fundamental}.

In \cite{li2018fundamental}, by formulating and characterizing a fundamental tradeoff between ``computation load" in the Map phase and ``communication load" in the Shuffle phase, Li  {\em et al.} demonstrated that these two quantities are inversely proportional to each other. This means that if each Map function is computed $r$ times, each of which is at a carefully chosen node, then the communication load in the Shuffle phase can be reduced by the same factor. This multiplicative gain in the Shuffle phase is achieved by the so-called ``coded distributed computing", which leverages the coding opportunities created in the Map phase by strategically placing the input files in all computing nodes. Note that one implicit assumption in this paper is that each computing node computes all possible intermediate values using their local files regardless whether these intermediate values will be used or not. In addition, there are two limitations of the proposed coded distributed computing scheme in \cite{li2018fundamental}. First, it requires an exponentially large number of input files when the number of computing nodes gets large. Second, it forces every $s$ computing nodes to compute one Map function, which leads to the requirement of a large number of Map functions, and hence a large number of output functions, in order to achieve the promised gain.

Some other aspects of coded distributed computing have been investigated in the literature. In \cite{ezzeldin2017communication}, Ezzeldin {\em et al.} revisited the computation-communication tradeoff by computing only necessary intermediate values in each node. A lower bound on the corresponding computation load was derived and a heuristic scheme, which  achieves the lower bound under some parameter regimes, was proposed.  In \cite{song2017benefit}, Song {\em et al.} considered the case where each computing node has access to a random subset of the input files and the system is asymmetric, which means that not all output functions depend on the entire data sets and we can decide which node computes which functions. The corresponding communication load was characterized. Interestingly, under some system parameters, no Shuffle phase is needed. In \cite{kiamari2017heterogeneous}, Kiamari {\em et al.} studied the scenario where different nodes can have different storage or computing capabilities. The proposed achievable scheme achieves the information-theoretical optimality of the minimum communication load in a system of $3$ nodes.

{\em Contributions:} In this paper, we consider the similar system configuration as in \cite{li2018fundamental} and propose a novel coded distributed computing approach based on a combinatorial design, which addresses the two limitations of the %coded distributed computing
scheme proposed in \cite{li2018fundamental} as follows. First, the proposed approach requires an exponentially less number of input files compared to that in \cite{li2018fundamental} for large $r$. Second, the required number of Map functions is also reduced exponentially when $s$ goes large. Meanwhile, the resulting computation-communication trade-off maintains the multiplicative gain compared to conventional uncoded unicast and is close to the optimal trade-off proposed in \cite{li2018fundamental}. In addition, our proposed scheme achieves the information theoretic lower bound asymmetrically for some system parameters.
%Due to the page limit, in the following, we omit the proofs which can be found in \cite{woolsey2017computing}.

\section{Network Model and Problem Formulation}
\label{sec: Network Model and Problem Formulation}

The network model is adopted from \cite{li2018fundamental}. We consider a distributed computing network where a set of $K$ nodes, labeled as $\{1, \ldots , K \}$, has the goal of computing $Q$ output functions and computing any one function requires access to $N$ input files. The $N$ input files, $\{w_1 , \ldots , w_N \}$, are assumed to be of equal size $B$ bits. The set of $Q$ output functions is $\{ \phi_1 , \ldots \phi_Q\}$ and each node $k\in \{ 1, \ldots , K \} $ is assigned to compute a set of output functions. We define $\mathcal{W}_k \subset \{ 1, \ldots Q \} $ as the indices of the output functions node $k$ is responsible for computing. The result of output function $i \in \{1, \ldots Q \}$ is $u_i = \phi_i \left( w_1, \ldots , w_N \right)$.

Alternatively, the output function can be computed by use of ``Map" and ``Reduce" functions such that $u_i = h_i \left( g_{i,1}\left( w_1 \right), \ldots , g_{i,N}\left( w_N \right) \right)$ where for every output function $i$ there exists a set of $N$ Map functions $\{ g_{i,1}, \ldots , g_{i,N}\}$ and one Reduce function $h_i$. Furthermore, we define the output of the Map function, $v_{i,j}=g_{i,j}\left( w_j \right)$, as the intermediate value resulting from performing the Map function for output function $i$ on files $w_j$. There are a total of $QN$ intermediate values and we assume that each has a length of $T$ bits.

The MapReduce distributed computing structure allows nodes to compute output functions without having access to all $N$ files. Instead, each node has access to $M$ out of the $N$ files and we define the set of files available to node $k$ as $\mathcal{M}_k \subset \{ w_1, \ldots , w_N\}$. Collectively the nodes use the Map functions to compute every intermediate value in the {\em Map} phase. Then, in the {\em Shuffle} phase, nodes multicast the computed intermediate values amongst one another via a shared link. %wireless channel.
The Shuffle phase is necessary so that each node can receive necessary intermediate values that it could not compute itself. Finally, in the {\em Reduce} phase, nodes use the Reduce functions with the appropriate intermediate values as inputs to compute the assigned output functions. %they are responsible for.

This distributed computing network designs yield two important parameters: the computation load, $r$, and the communication load, $L$. The computation load is defined as the average number of times each intermediate value is computed among all nodes. In other words, the computation load is the number of intermediate values computed in the Map phase normalized by the total number of unique intermediate values, $QN$. The communication load is defined as the amount of traffic load (in bits) among all nodes in the Shuffle phase %channel use in bits required for the Shuffle phase
normalized by $QNT$. We define the computation-communication function as
\be
L^*(r) \eqdef \inf\{L: (r, L) \text{ is feasible}\}.
\ee
Throughout this paper, we consider a few different design options. We enforce one of the following assumptions: 1) each computing node only computes the necessary intermediate values as in \cite{ezzeldin2017communication}; %the number of intermediate values computed from a given file linearly contribute to the computation load or
2) each computing node computes all possible intermediate values. %computing any number of intermediate values for a given file contributes %the contributes
%the same amount to the computation load.\footnote{WordCount is a classic example where counting the occurrence of one word in the text file requires roughly the same amount of computation as counting the occurrence of two or more words.}
%In the first scenario, nodes will strategically compute only necessary intermediate values for use in transmitting, decoding or computing a Reduce function. The the second scenario, nodes will compute every intermediate value for a given file.
Furthermore, in this paper, we first discuss the design scenario when all of the $Q$ output functions are computed exactly once and $|\mathcal{W}_i \cap \mathcal{W}_j| = 0$ for $i\neq j$. We then expand our model such that output functions are computed at multiple nodes. We define $s$ as the number of nodes which calculate each output function.

%We enforce one of the following assumptions: %1)
% the number of intermediate values computed from a given file linearly contribute to the computation load,  %or 2) computing any number of intermediate values for a given file contributes %the contributes
%%the same amount to the computation load.\footnote{WordCount is a classic example where counting the occurrence of one word in the text file requires roughly the same amount of computation as counting the occurrence of two or more words.} In the first scenario,
%meaning that nodes will strategically compute only necessary intermediate values for use in transmitting, decoding or computing a Reduce function. %The the second scenario, nodes will compute every intermediate value for a given file.
%Furthermore, in this paper, we first discuss the design scenario when all of the $Q$ output functions are computed exactly once and $|\mathcal{W}_i \cap \mathcal{W}_j| = 0$ for $i\neq j$. We then expand our model such that output functions are computed at multiple nodes. We define $s$ as the number of nodes which calculate each output function.

\section{Hypercube Computing Approach ($s=1$)}
\label{sec: Hypercube Caching Network Approach}

In this section, we consider the case when $s=1$. Every output function is calculated exactly once in the Reduce phase. %To have an achievable scheme we find that
%For mathematical convenience,
Let $Q=\eta_2 K$, where $\eta_2 \in \mathbb{Z}^+$, such that every node $k$ computes a set $\eta_2$ distinct functions. We construct a hypercube lattice of dimension $d$ with the length of each side $x$ to describe the file availability among all nodes (see Fig. \ref{fig: 3d fig}). Each lattice point represents a set of $\eta_1$ files, where $\eta_1 \in \mathbb{Z}^+$, and nodes have a set of files available to them represented by a ``hyperplane" of lattice points. %In the case where
When $s=1$, we show that the requirement on the number of files is $N = \eta_1 x^d$.  We first present an example when $d=3$. In this case, the ``hypercube" becomes a ``cube" (see Fig. \ref{fig: 3d fig}). %where $d=2$.

\subsection{An Example (3-Dimension)}
\label{sec: 3D_1}
\begin{figure}
\centering
\centering \includegraphics[width=7cm]{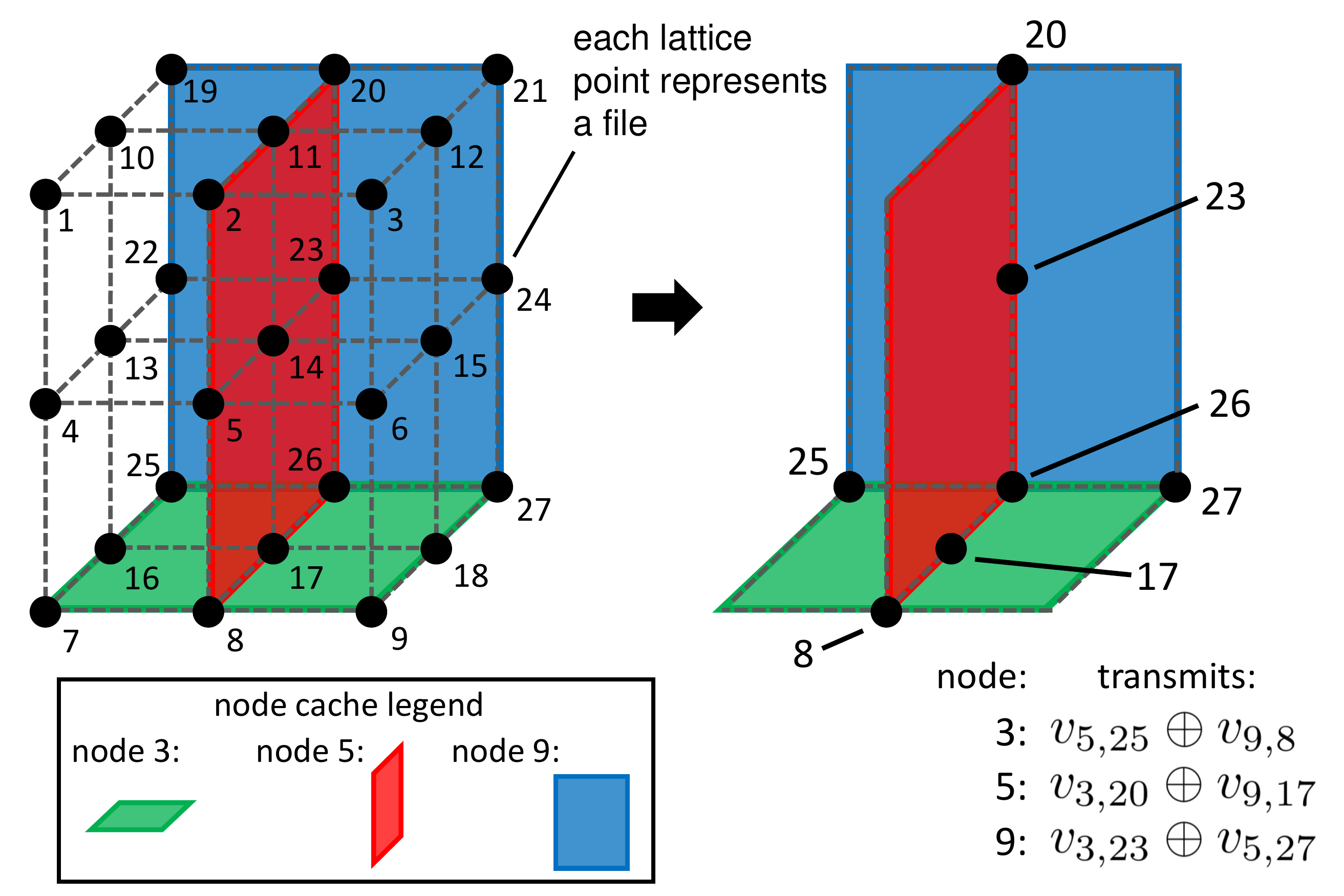} %width=9cm, height=6cm
\vspace{-0.2cm}
\caption{~\small Depiction of the hypercube scheme where $d=3$ and $x=3$. Each lattice point represents a file and each of the $K=9$ nodes has set of files available to it represented by a plane of lattice points. The green, red and blue planes represent the files locally available to nodes 3, 5 and 9, respectively. The intersections of these planes represent files that multiple nodes have available which yields coded multicasting opportunities.}
\label{fig: 3d fig}
\vspace{-0.4cm}
\end{figure}

%To demonstrate the general scheme
To illustrate our idea, we consider a ($d=3$)-dimensional hypercube (cube in this case) where each dimension is $x=3$ lattice points in length. There are $K=xd=9$ nodes and let node $i$ compute the function $i$, $\mathcal{W}_i=\{ i\}$. There are a total of $Q=9$ functions ($\eta_2=1$), each of which is computed exactly once. % by the network.
The $9$ computing nodes are split into $d=3$ groups of size $x=3$ where $\mathcal{K}_1 = \{1,2,3\}$, $\mathcal{K}_2 = \{4,5,6 \}$ and $\mathcal{K}_3 = \{7,8,9 \}$. We consider $N=27$ ($\eta_1=1$) files where each file is locally available to exactly $3$ nodes, one node from each set. This is analogous to defining a point in a 3-dimensional space for which the value of each dimension defines a node that has that file locally available as shown in Fig.~\ref{fig: 3d fig}. For example, the files locally available to nodes 3, 5 and 9 ($\mathcal{M}_3$, $\mathcal{M}_5$, and $\mathcal{M}_9$) are depicted by the green, red and blue planes respectively. %Essentially,
In fact, fixing one dimension and varying the rest dimensions define a set of files available to a node.
%The file set $\mathcal{B}_{26}$, which is simply the file $w_{26}$ in this case, is locally available to nodes $\{ 3, 5,9 \}=\mathcal{T}_{26}$.
Since $w_{26}$ is locally available to nodes $\{ 3, 5,9 \}$, we denote the node set $\{ 3, 5,9 \}$ as $\mathcal{T}_{26}$.

In the Map phase, each node computes the intermediate values from locally available files for the function it needs to compute. For example, node $5$ has the file set $\mathcal{M}_5=\{ w_2, w_5, w_8, w_{11}, w_{14}, w_{17}, w_{20}, w_{23}, w_{26}\}$ and will compute $\{ v_{5,j} : w_j\in \mathcal{M}_5 \}$ for use in calculation of function output $u_5$. Furthermore, for every $w_j\in \mathcal{M}_5$, node $5$ computes every intermediate value, $v_{i,j}$, such that $i\notin \mathcal{K}_2$ and node $i$ does not have $w_j$ locally available, $w_j\notin \mathcal{M}_i$. The motivation for this criteria is, first, node $5$ does not form multicasting groups with other nodes (nodes $4$ and $6$) aligned along the same dimension;\footnote{In Fig. \ref{fig: 3d fig}, node $4$ and $6$ caches the files from the two planes parallel to the red plane, which represents the files cached by node $5$.} second, there is no need to compute an intermediate value for a node that can compute it itself. Every other node computes intermediate values based on a similar approach.

We use the example of node set $\mathcal{T}_{26}$ to explain the Shuffle phase. We consider the $3$ sets of intermediate values for which each intermediate value is requested by $1$ node and computed by the other $2$ nodes. Those sets are $\mathcal{V}_{\{5,9\}}^{\{3\}} = \{ v_{3,20},v_{3,23} \}$, $\mathcal{V}_{\{3,9\}}^{\{5\}} = \{ v_{5,25},v_{5,27} \}$ and $\mathcal{V}_{\{3,5\}}^{\{9\}} = \{ v_{9,8},v_{9,17} \}$. Each set is split into $d-1=2$ subsets which get labeled based on which node transmits the subset. The subsets are $\mathcal{V}_{\{5,9\}}^{\{3\},5} = \{v_{3,20}\}$, $\mathcal{V}_{\{5,9\}}^{\{3\},9} = \{v_{3,23}\}$, $\mathcal{V}_{\{3,9\}}^{\{5\},3} = \{v_{5,25}\}$, $\mathcal{V}_{\{3,9\}}^{\{5\},9} = \{v_{5,27}\}$, $\mathcal{V}_{\{3,5\}}^{\{9\},3} = \{ v_{9,8} \}$ and $\mathcal{V}_{\{3,5\}}^{\{9\},5} = \{ v_{9,17} \}$. Nodes 3, 5 and 9 then collectively transmit these subsets in coded multicasts as described by (\ref{eq: 1_trans_eq}) in Section \ref{sec: genscheme_s1} and shown in Fig.~\ref{fig: 3d fig}. It is clear that each node can recover requested intermediate values from received coded multicasts as it has computed the other intermediate values of the multicast in the Map phase.

In this example, each node computes $9$ intermediate values for itself and requests $18$ intermediate values in order to compute the output function that it is responsible for. Each node is involved in $9$ multicasting groups for which it receives $2$ intermediate values from each group which satisfies all of its requests. In total, each node computes $45$ intermediate values consisting of $9$ for itself and $36$ to either transmit or decode received coded multicasts. Accounting for all nodes it is clear that $r_{\rm hc}= \frac{45 \cdot 9}{27 \cdot 9} =\frac{5}{3}$.\footnote{If each node compute all possible intermediate value, $r=d=3$, which is also the case considered in \cite{li2018fundamental}.} Also, each node transmits $9$ coded messages with the equivalent size of $1$ intermediate value each and therefore the communication load is $L_{\rm hc} = \frac{9 \cdot 9}{27} = \frac{1}{3}$. This demonstrates a significant decrease compared to the uncoded communication load, $L_{\rm uncoded}=1-\frac{M}{N}=\frac{2}{3}$. If each node can compute all possible intermediate values and by using the approach proposed in \cite{li2018fundamental}, we can obtain $L^* = \frac{2}{9}$, which is slightly lower than $L_{\rm hc}$. However, the minimum number of files needed is ${9 \choose 3} = 84$.

\subsection{General Scheme for $s=1$}
\label{sec: genscheme_s1}
For a network of $K$ nodes which collectively computes $Q$ functions exactly once, each node $k\in\{1,\ldots K\}$ computes a set of functions, $\mathcal{W}_k\subset \{1,\dots ,Q\}$, such that $|\mathcal{W}_k|=\eta_2$, $|\mathcal{W}_i \cap \mathcal{W}_j|=0$ when $i\neq j$, $Q=K\eta_2$ and $K,Q,\eta_2\in \mathbb{Z}^+$. Nodes are split into $d$ disjoint sets each of $x$ nodes denoted by $\{\mathcal{K}_1,\ldots ,\mathcal{K}_d\}$ where $K=xd$ where $x,d\in \mathbb{Z}^+$. To define file availability, consider all node sets $\{\mathcal{T}:|\mathcal{T}\cap\mathcal{K}_i|=1\text{ }\forall \text{ } i\in \{1,\ldots , d\}\}$. There are $x^d$ such sets which will be denoted as $\{\mathcal{T}_1,\ldots ,\mathcal{T}_{x^d}\}$. The $N$ files are split into $x^d$ disjoint sets labeled as $\{\mathcal{B}_{1},\ldots,\mathcal{B}_{{x^d}}\}$ and file set $\mathcal{B}_{i}$ is available only to nodes of set $\mathcal{T}_i$. These file sets are of size $\eta_1\in \mathbb{Z}^+$ and $N=x^d\eta_1$. Furthermore, define $\mathcal{M}_k=\{\mathcal{B}_i : k\in \mathcal{T}_i \}$, which is the set of files available to node $k$. The Map, Shuffle and Reduce phases are defined as follows:

\begin{itemize}
%\item {\bf Map Phase:} Each node $k\in\{1,\ldots K\}$ computes: 1) every intermediate value, $v_{i,j}$, such that $i\in \mathcal{W}_k$ and $w_j\in \mathcal{M}_k$ and 2) every intermediate value, $v_{i,j}$, such that $w_j\in \mathcal{B}_\ell \subset \mathcal{M}_k$ and $i\in \{\mathcal{W}_z : z\in\{1,\ldots ,K \}\setminus \left(\mathcal{T}_\ell \cup \mathcal{K}_p \right)\}$ where $k\in \mathcal{K}_p$.
\item {\bf Map Phase:} Each node $k\in\{1,\ldots K\}$ computes: 1) every intermediate value, $v_{i,j}$, such that $i\in \mathcal{W}_k$ and $w_j\in \mathcal{M}_k$ and 2) every intermediate value, $v_{i,j}$, such that $w_j\in \mathcal{M}_k$ and $i\in \{\mathcal{W}_z : w_j \notin \mathcal{M}_z, z\in\{1,\ldots ,K \}\setminus \mathcal{K}_p\}$ where %$k\in \mathcal{K}_p$.
$\Kc_p = \{\Kc_i: k \in \Kc_i, i \in \{1, \cdots, d\}\}$.

\item {\bf Shuffle Phase:} For all $n\in \{1,\ldots ,x^d\}$ and for all nodes $z\in \mathcal{T}_n$ consider the set of intermediate values
    \be
    \mathcal{V}_{\mathcal{T}_n\setminus z}^{\{z\}}=\{ v_{i,j}: i\in \mathcal{W}_z,\text{ }w_j\in \bigcap \limits_{k\in\mathcal{T}_n\setminus z}\mathcal{M}_k \}
    \ee
     which are requested only by node $z$ and computed at nodes $\mathcal{T}_n\setminus z$. Furthermore, $\mathcal{V}_{\mathcal{T}_n\setminus z}^{\{z\}}$ is split into $d-1$ disjoint sets of equal size denoted by $\{ \mathcal{V}_{\mathcal{T}_n\setminus z}^{\{z\},\sigma_1},\ldots , \mathcal{V}_{\mathcal{T}_n\setminus z}^{\{z\},\sigma_{d-1}} \}=\mathcal{V}_{\mathcal{T}_n\setminus z}^{\{z\}}$ where $\{ \sigma_1,\ldots \sigma_{d-1}\}=\mathcal{T}_n\setminus z$. Each node $k\in \mathcal{T}_n$ multicasts
    \be
    \label{eq: 1_trans_eq}
    \bigoplus \limits_{z\in \mathcal{T}_n\setminus k} \mathcal{V}_{\mathcal{T}_n\setminus z}^{\{z\},k}
    \ee
%$\mathcal{V}_{\mathcal{T}_n\setminus z}^{\{z\}}=\{ v_{i,j}: i\in \mathcal{W}_z,\text{ }w_j\in \mathcal{M}_k,\text{ }k\in\mathcal{T}_n\setminus z \}$
\item {\bf Reduce Phase:} For all $k\in \{ 1, \ldots K\}$, node $k$ computes all output values $u_q$ such that $q\in \mathcal{W}_k$.

%collectively transmit every intermediate value, $v_{i,j}$, such that $i\in \mathcal{W}_z$, $z\in \mathcal{S}_n$ and $w_j\in\{\mathcal{M}_k : k \in \mathcal{S}_n\setminus k \}$. More specifically,
\end{itemize}

\subsection{Achievable Computation and Communication Load}

In this section we derive the computation and communication load for the proposed scheme utilizing a hypercube with an arbitrary number of dimensions and size of each dimension. We derive these values in two scenarios. First, as has been discussed so far, we consider the case when nodes compute a subset of the $Q$ intermediate values %sub-functions
for any given file and only necessary intermediate values are computed. %the contribution to the computation load is linear with respect to the number of sub-functions computed. %In some cases this is unrealistic and
In addition, we also consider the case when a node will compute all possible intermediate values %sub-functions for every available file
and demonstrate that a small modification to the scheme of section \ref{sec: genscheme_s1} accommodates this assumption.

The following theorem evaluates the computation and communication load for the hypercube scheme when only necessary intermediate values are computed. %the number of sub-functions calculates per file contributes linearly to the computation load.

\begin{theorem}
\label{theorem: 1}
Let $K,Q,N,M$ be the number of nodes, number of functions, number of files and number of files available to each node, respectively. For some $x,d,\eta_1,\eta_2 \in \mathbb{Z}^+$ such that $d\geq 2$, $K=xd$, $N = \eta_1 x^d $ and $Q = \eta_2 x d $, the following computation and communication load pair is achievable:
\be
\label{eq: theorem 1}
r_{\rm hc} = \frac{1+\left( d-1 \right) \left( x-1 \right)}{x},
\ee
\be
\label{eq: theorem 1}
L_{\rm hc} = \frac{x-1}{x\left( d-1 \right)}.
\ee
\hfill $\square$
\end{theorem}
From this theorem, we can observe that when $x \gg 1$, %or equivalently $K \gg d$, that $r_{\rm hc} \approx d-1$ and
$L_{\rm hc} \approx \frac{1}{d-1} \approx \frac{1}{r_{\rm hc}}$. %. Therefore, when $K \gg r_{\rm hc}$
This means that the communication load is inversely proportional to the computation load, which is also shown in \cite{li2018fundamental}.

We extend this scheme to accommodate the assumption that each node computes all possible intermediate values. %computing intermediate values from a particular file does not linearly contribute to the computation load. In other words, a node performs every sub-function on every file it has available to it.
%The following describes the achievable computation and communication load in this scenario.
\begin{corollary}
\label{corollary: 1}
Let $K,Q,N,M$ be the number of nodes, number of functions, number of files and number of files available to each node, respectively. For some $x,d,\eta_1,\eta_2 \in \mathbb{Z}^+$ such that $d\geq 2$, $K=xd$, $N = \eta_1 x^d $, $Q = \eta_2 x d $ and assuming every node computes all possible intermediate values from available files, the following computation and communication load pair is achievable:
\be
\label{eq: theorem 2}
r_{\rm hc}' = d,
\ee
\be
\label{eq: theorem 2}
L_{\rm hc}' = L_{\rm hc} = \frac{x-1}{x\left( d-1 \right)}.
\ee
\hfill$\square$
\end{corollary}
%\begin{IEEEproof}
%We use a modified version of the Map phase in section \ref{sec: genscheme_s1} which is: every node $k\in \{1,\ldots ,K \}$ computes the set of intermediate values $\{v_{i,j} : i\in \{ 1, \ldots , Q\}, w_j \in \mathcal{M}_k \}$. Every file is available to set of $d$ nodes, therefore, every intermediate value is computed $d$ times and $r_{\rm hc}'=d$. By recognizing that in the modified Map phase every node computes every intermediate that it would have in the unmodified Map phase, it is clear that Shuffle phase of section \ref{sec: genscheme_s1} is still certainly possible. By performing the Shuffle phase, we yield the same communication load as we would if the unmodified Map phase was used, therefore, $L_{\rm hc}'=L_{\rm hc}=\frac{x-1}{x\left( d-1 \right)}$.
%\end{IEEEproof}
%The relationship between $L_{\rm hc}'$ and $r_{\rm hc}'$ is made clear by substituting $d = r_{\rm hc}'$ and $x = \frac{K}{d} = \frac{K}{r_{\rm hc}'}$ to obtain
By using Corollary \ref{corollary: 1}, we obtain
%\be
%\label{eq: comp_s1}
$L_{\rm hc}' %= %\frac{\frac{K}{r_{\rm hc}'}-1}{\frac{K}{r_{\rm hc}'}\left( r_{\rm hc}' - 1 \right)}
%= \frac{K-r_{\rm hc}'}{K\left( r_{\rm hc}'-1 \right)}
= \frac{1-\frac{r_{\rm hc}'}{K}}{r_{\rm hc}'-1}.$
%\ee

\section{Hypercube Computing Approach ($s=d$)}

The work in this section is motivated by the fact that distributed computing systems generally perform multiple rounds of Map Reduce computations. The results from the $Q$ output functions become the input files for the next round. To have consecutive Map Reduce algorithms which take advantage of the computation-communication load trade-off, it is important that each function is computed at multiple nodes. We define $s$ as the number of times each function is computed. Alternatively, $s$ can be defined by the number of nodes which compute any given output function. To implement consecutive rounds of Map Reduce using the hypercube method, we construct the network by using hypercube not only to define the input files that each node has, %nodes have available to them,
but also to define the output functions each node is responsible for computing. Thus, in the following Map Reduce round, the hypercube approach can be used again. In this section, we describe how to the use the hypercube computing approach such that $s=d$.

\subsection{An Example (2-Dimensional)}
\label{sec: 2d_sgt1}
\begin{figure}
\centering
\centering \includegraphics[width=7cm]{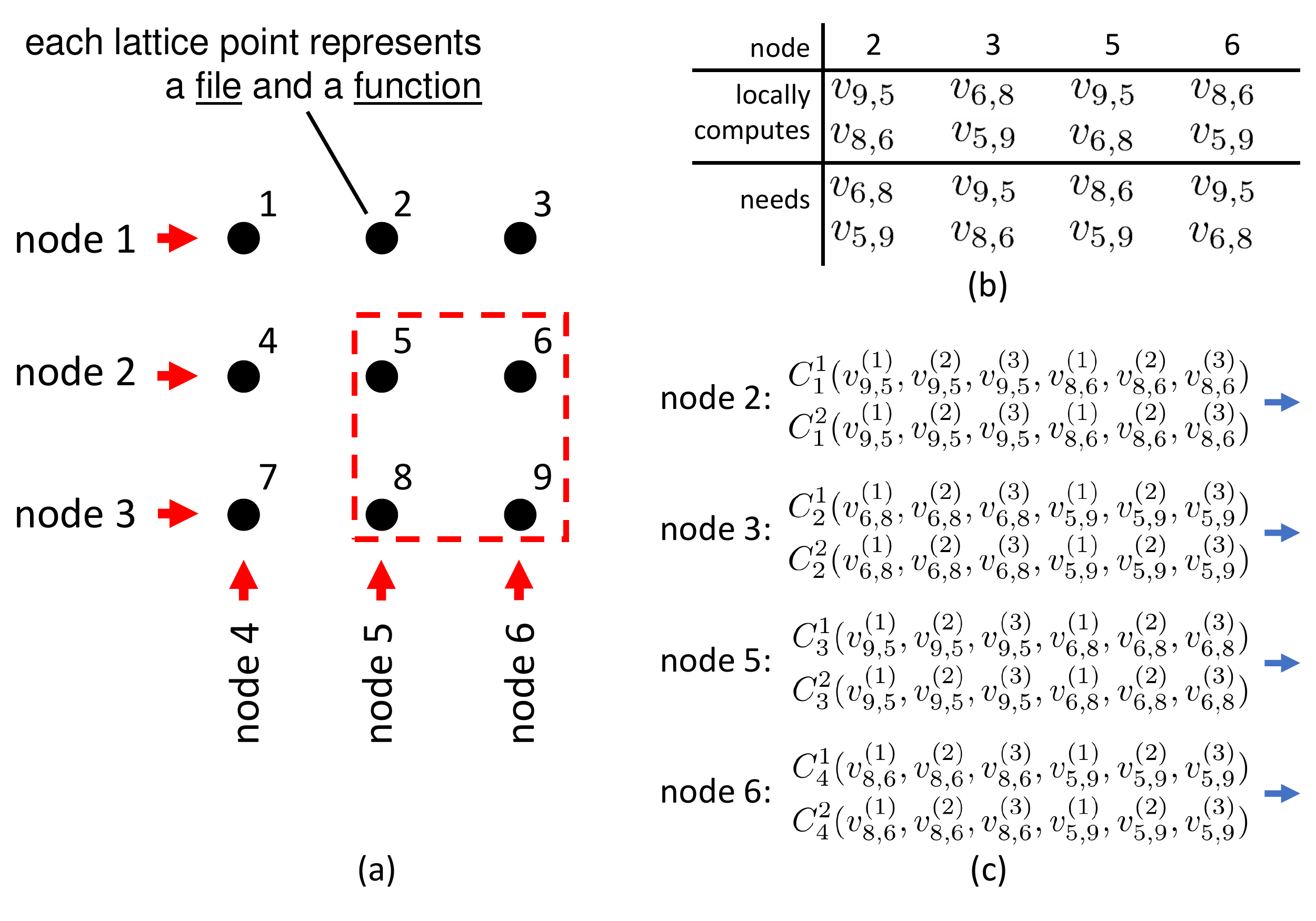} %width=9cm, height=6.3cm
\vspace{-0.2cm}
\caption{~\small An example of the proposed scheme in a distributed computing network with $K=6$, $s=d=2$, and $N=Q=9$. %$n=4$, $M/N=1/2$, $N=4\eta_1$ and $Q=4\eta_2$.
Each point on the lattice represents both a file and a computing function. Each node has a set of files %available to it
represented by a line. $C_i^j(\cdot)$ represent random independent linear combinations. %$f_i$ represents the $\eta_2$ functions computed by node $i$.
}
\label{fig: 2d fig_sgt1}
\vspace{-0.4cm}
\end{figure}

For simplicity, we consider a ($d=2$)-dimensional hypercube (plane) where each side is $x=3$ lattice points in length. There are $K=6$ nodes and nodes $1$, $2$ and $3$ are aligned along one dimension and nodes $4$, $5$ and $6$ are aligned along the other dimension as shown in Fig.~\ref{fig: 2d fig_sgt1}(a). Each lattice point represents an input file as well as an output function. The files available to a node and the output functions computed by a node are determined by a line of lattice points that a node is aligned with. For example, node $1$ has the input files $w_1$, $w_2$ and $w_3$ %locally available to it
and node $1$ is responsible for computing the output functions $\phi_1$, $\phi_2$ and $\phi_3$. While $\phi_1$ is computed at both nodes $1$ and $4$, %Similarly, input files $w_2$, $w_5$ and $w_8$ are locally available to node 5 and output functions $\phi_2$, $\phi_5$ and $\phi_8$ are computed by node 5.
it can be seen that similar to the case when $s=1$ where files are assigned to $d$ nodes based on the hypercube lattice, now output functions are assigned to $d$ nodes. In this example, $s=d=2$, which is the number of dimensions in the hypercube.

Note that since different intermediate values may be computed different times, the Map and Shuffle phases can be defined by three rounds where nodes compute intermediate values requested by $0$, $1$, or $2$ nodes, respectively. We define an intermediate value requested by $0$ nodes as an intermediate value for which the nodes that need it can compute it locally. %to compute an output function also have the file locally available to compute that intermediate value.
For example, $v_{2,2}$ is considered an intermediate value which is requested by $0$ nodes. Nodes $1$ and $5$ are the only nodes that need to compute $u_2$ and need $v_{2,2}$. However, it is clear that both nodes $1$ and $5$ also have access to the file $w_2$ and can compute this intermediate value themselves. Hence, $v_{2,2}$ does not need to be transmitted in the Shuffle phase. In this example, any intermediate value $v_{i,i}$ for $i\in \{1, \dots, 9 \}$ is an intermediate value requested by $0$ nodes.

Next, we consider intermediate values requested by a $1$ node such as $v_{2,5}$ and $v_{2,8}$. Nodes $1$ and $5$ are the only nodes that need $v_{2,5}$ and $v_{2,8}$, However, node $5$ can compute these intermediate values itself, while node $1$ does not have access to $w_5$ and $w_8$. The opposite is true of intermediate values $v_{2,1}$ and $v_{2,3}$. Nodes $1$ and $5$ can unicast these intermediate values to each other in the Shuffle phase. All of the intermediate values requested by $1$ node can be found by considering all pairs of nodes such that there is $1$ node aligned along each dimension. For each pair, there exists an output function that is computed by the nodes of this pair and not computed by any other nodes. Furthermore, each node has access to $2$ files that the other node does not, therefore, each node of the pair computes $2$ intermediate values that are only requested by the other node of the pair.

In the last round, we consider intermediate values which are requested by $2$ nodes such as $v_{5,9}$. Both nodes $2$ and $5$ compute output function $5$ ($u_5$). However, neither has access to $w_9$. We can recognize that $w_9$ is available to two nodes which are nodes $3$ and $6$. Importantly, we also observe that both nodes $3$ and $6$ request $v_{9,5}$ which can be computed at nodes $2$ and $5$. Among these four nodes we also see that $v_{6,8}$ is requested by nodes $2$ and $6$ and can be computed by nodes $3$ and $5$ and the opposite is true for $v_{8,6}$. These observations are summarized in Fig.~\ref{fig: 2d fig_sgt1}(b) and the lattice points which represent these input files and input functions are highlighted in Fig.~\ref{fig: 2d fig_sgt1}(a). %There are multiple ways for these nodes to transmit these intermediate values amongst each other. To optimally reduce the impact to the communication load,
In order to transmit, each intermediate value can be split into $3$ packets and each node requests $6$ %of {\RED a word missing?}
packets and has the other $6$ locally computed. Each node transmits $2$ random linear combinations of its locally available packets as shown in Fig.~\ref{fig: 2d fig_sgt1}(c). As a result, each node will receive $6$ equations, together with what it has it can solve for the $6$ unknowns. Overall, this round consists of considering all groups of $4$ nodes such that there are $2$ nodes aligned along each dimension.

In this example, we observe that $Q=N=x^d=9$ and therefore the total number of unique intermediate values is $QN=x^{2d}=81$. It can be computed that $r_{\rm hc} = \frac{14}{9}$ and $L_{\rm hc} = \frac{20}{27}$. While if each user computes all possible intermediate values, by using the approach in \cite{li2018fundamental}, we obtain that $L^* = \frac{8}{15}$. However, our approach only requires $N=9$ files and $Q=9$ functions, while the approach in \cite{li2018fundamental} requires $N=15$ files and $Q=15$ functions.

\subsection{Achievable Computation and Communication Load}

The following theorem evaluates the computation and communication load for the hypercube scheme when only necessary intermediate values are computed. %the number of intermediate value %sub-functions
%calculated per file contributes linearly to the computation load. %Using the similar approach of Corollary 1, it can be observed that when each node computes all possible interemdiate values, %The case when each node computes all possible intermediate values can be found in \cite{woolsey2017computing}.\footnote{It can also be observed that when each node computes all possible intermediate values, $r_{\rm hc} = d$ and }

\begin{theorem}
\label{theorem: 2}
Let $K,Q,N,M,s$ be the number of nodes, number of functions, number of files, number of files available to each node, and number of nodes which compute each function, respectively. For some $x,d,\eta_1,\eta_2 \in \mathbb{Z}^+$ such that $d\geq 2$, $s=d$, $K=xd$, $N = \eta_1 x^d $ and $Q = \eta_2 x d $, the following computation and communication load pair is achievable:
\be
\label{eq: theorem 21}
r_{\rm hc} = \frac{d\left( x^d - x + 1 \right)}{x^d},
\ee
\be
\label{eq: theorem 2}
L_{\rm hc} = \frac{ d\left( x-1 \right) }{x^d\left( d-1 \right)} + \frac{1}{x^{2d}}\sum_{\gamma = 2}^{d} 2{d \choose \gamma}{x \choose 2}^\gamma x^{d-\gamma}\gamma\frac{2^{\left( \gamma - 1\right)}}{2\gamma - 1}.
\ee
%{\RED need new labeling for r and L?}
\hfill$\square$
\end{theorem}
Using the similar approach of Corollary 1, it can be observed that when each node computes all possible intermediate values, (\ref{eq: theorem 21}) becomes  $r_{\rm hc} = d$ and (\ref{eq: theorem 2}) stays unchanged.

\section{Performance Analysis and Discussion}

In order to do the fair comparison, we assume that each node computes all possible intermediate values.

\subsection{The requirement of $N$ and $Q$}

%Under the assumption that each node computes all the possible intermediate values. It is possible to compare the
In this section, our goal is to compute the minimum required $N$ and $Q$ of the proposed scheme and the optimal scheme in \cite{li2018fundamental}. By our construction, it can be seen that the minimum requirements of $N$ and $Q$ are $\left(\frac{K}{r}\right)^r$ and $\left(\frac{K}{s}\right)^s$ respectively. %By using the approach proposed in \cite{li2018fundamental},
While the minimum requirements of $N$ and $Q$ in \cite{li2018fundamental} are ${K \choose r}$ and ${K \choose s}$. Hence, it can be observed that the proposed approach reduces the required numbers of both $N$ and $Q$ exponentially as a function of $r$ and $s$.

\subsection{Optimality}
Although the required $N$ and $Q$ are reduced significantly, we can still guarantee the performance of the proposed approach in terms of computation-communication function.
%One of our goals is designing this new computing approach was to reduce the constraints on the necessary number of input files and output functions while maintaining a near optimal computation-communication load trade off.
If each node computes all possible intermediate values, then optimal computation-communication function is given by \cite{li2018fundamental}
\be
\label{eq: communication load optimal}
L^*\left( r,s \right) = \sum_{\ell = \max \{ r+1, s\}}^{\min \{ r+s, K \}} \frac{\ell {K \choose \ell} {\ell-2 \choose r-1}{r \choose \ell - s}}{r {K \choose r}{K \choose s}}.
\ee
When $s=1$, the optimality of the proposed approach is given by the following corollary.
\begin{corollary}
When $s=1$ and under the assumption in Theorem \ref{theorem: 1}, $L_{\rm hc}$ achieves information theoretic optimality when $r \rightarrow \infty$.
\hfill$\square$
\end{corollary}
%\begin{IEEEproof}
%%We compare our results to the coded distributed computing design in \cite{li2018fundamental} which was proven to be optimal assuming every node computes all possible intermediate values given the available input files. Te authors demonstrated that for $N_1 = {K \choose r}\eta_1$ input files and $Q_1 = {K \choose s}\eta_2$ output functions that the following communication load is possible
%%\be
%%\label{eq: communication load optimal}
%%L_1'\left( r,s \right) = \sum_{\ell = \max \{ r+1, s\}}^{\min \{ r+s, K \}} \frac{\ell {K \choose \ell} {\ell-2 \choose r-1}{r \choose \ell - s}}{r {K \choose r}{K \choose s}}.
%%\ee
%%First, we compare the results when $s=1$ and it is clear that the result from \cite{li2018fundamental} simplify to
%When $s=1$, (\ref{eq: communication load optimal}) becomes
%\be
%L^*\left( r,1 \right) = \frac{K-r}{rK} = \frac{1}{r}\left( 1 - \frac{r}{K} \right).
%\ee
%Using (\ref{eq: comp_s1}), we obtain %the following comparison
%\be
%\frac{L_{\rm hc}'}{L^*\left( r,1 \right)} = \frac{rK}{K-r} \frac{K-r}{K\left( r-1 \right)} = \frac{r}{r-1}
%\ee
%which goes to $1$ as $r \rightarrow \infty$. %where we subsistence $r=r_{\rm hc}'$ for the purposes of comparison.
%\end{IEEEproof}

\begin{figure}

\centering
%\subfigure[]{
\centering \includegraphics[width=7cm]{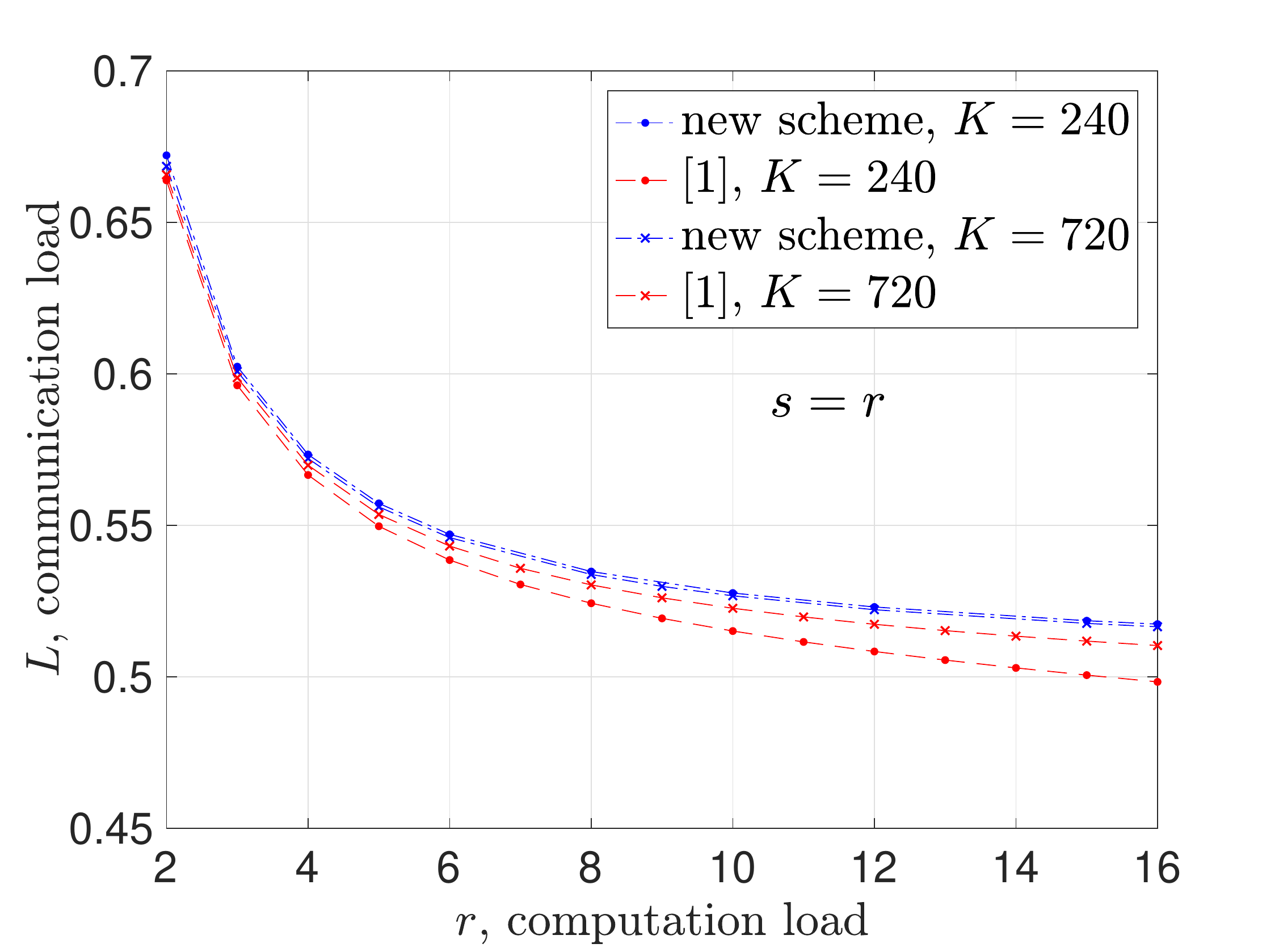} %width=9cm, height=7cm
\vspace{-0.3cm}
\caption{~\small A comparison of the communication load by the proposed and the state-of-the-art distributed computing schemes.}
\label{fig: srgraph}
\vspace{-0.4cm}
\end{figure}

When $s\geq2$, the relation between (\ref{eq: communication load optimal}) and (\ref{eq: theorem 2}) is not obvious due to their complicated formats. From Fig. \ref{fig: srgraph}, it can be observed that first the communication load of the proposed scheme has a multiplicative gain compared to conventional uncoded unicast ($L(r) = (1-\frac{r}{K})$); second, the communication load of the proposed scheme is close to that in \cite{li2018fundamental} especially when $s=r$ is relatively small. In addition, we can also prove the asymptotic optimality of the proposed scheme when $s=2$.
\begin{corollary}
When $s=2$ and under the assumption in Theorem \ref{theorem: 2}, $L_{\rm hc}$ achieves information theoretic optimality when $K \rightarrow \infty$. \hfill$\square$
\end{corollary}

\appendices
\bibliographystyle{IEEEbib}
\bibliography{references_d2d}

\end{document}